\documentclass[fleqn,usenatbib]{mnras}

\usepackage{bm}

\usepackage{newtxtext,newtxmath}
\usepackage[T1]{fontenc}
\usepackage{tabularx}
\DeclareRobustCommand{\VAN}[3]{#2}
\let\VANthebibliography\thebibliography
\def\thebibliography{\DeclareRobustCommand{\VAN}[3]{##3}\VANthebibliography}

\usepackage{bm}

\usepackage[dvipsnames]{xcolor} % for textcolor colors - SG

\usepackage{graphicx}	% Including figure files
\usepackage{amsmath}	% Advanced maths commands
\usepackage{multicol}
\title[Impact of ISM on black holes]{The effect of interstellar medium on LVK's black holes}

\author[Sohan Ghodla]{Sohan Ghodla$^{\thanks{sohanghodla9@gmail.com}{}}$ \\
Department of Physics, University of Auckland, Private Bag 92019, Auckland, New Zealand}

\date{Accepted XXX. Received YYY; in original form ZZZ}

\pubyear{2024}

\begin{document}
\label{firstpage}
\pagerange{\pageref{firstpage}--\pageref{lastpage}}
\maketitle


\begin{abstract}
   Gravitational radiation alone is not efficient in hardening the orbit of a wide binary black hole (BBH). By employing a toy model for the interstellar medium (ISM) surrounding BBHs, here we discuss the effect of this baryonic medium on BBH dynamics. Depending on the BBH's mass, we show that a binary surrounded by an isotropic cold neutral medium (i.e., an asymptotic temperature $T_{\infty} \approx 100$ K) with a time-averaged particle density of $\langle n_H \rangle = \mathcal{O}(1)$ cm$^{-3}$ can play a significant role in hardening the binary orbit over a $\mathcal{O}(10^9)$ yr time scale. Additionally, this causes the black hole's mass to grow at a rate $\propto m^2$. We thus discuss the impact of the ISM on the LIGO-Virgo-KAGRA (LVK) observables and quantify the properties of the ISM under which the latter could act as an additional important pathway for driving a subset of LVK's BBH mergers.
   
   % We find that under the condition that $\mathcal{O}(1)$\% of the BBHs are surrounded by uniform isotropic cold neutral medium with reasonable values of  $\langle n_H \rangle$, ISM could be a major 

\end{abstract}



%%%%%%%%%%%%%%%%%%%%%%%%%%%%%%%%%%%%%%%%%%%%%%%%%%

%%%%%%%%%%%%%%%%% BODY OF PAPER %%%%%%%%%%%%%%%%%%

\section{Introduction}

Emission of gravitational radiation alone is not efficient in hardening the orbit of a binary black hole (BBH). The merger time of a binary can be calculated entirely in the weak field limit, and for a circular BBH (with equal mass components) to merge within a Hubble time, the binary separation $r$ has to satisfy (cf. \citealt{Peters_1964} or Eq.~\ref{eq: Peters merger time} later)
\begin{equation}
    r \leq \left(\frac{512 G^3 m^3}{5 H_0 c^5} \right)^{1/4}  = 3.7 m^{3/4} \, R_\odot \,,
    \label{eq: seperation to merge in a Hubble time}
\end{equation}
where $H_0, m$ is the Hubble's constant and the mass of the component black hole (in  $M_\odot$), respectively. For a typically stellar mass black hole with $m \leq 50 M_\odot$, Eq.~\ref{eq: seperation to merge in a Hubble time} requires $r \lesssim 69 R_{\odot}$ which is much smaller than the radial extent of most stars during the later phase of their evolution. However, the LIGO-Virgo-KAGRA (LVK) network of ground-based gravitational wave detectors has already observed close to eighty BBH mergers \citep{Abbott_O3a_2021, Abbott_O3b_2021}.

Although multiple 1D binary stellar evolution studies are able to generate tight BBHs from progenitor binary stars with relatively large initial separation (e.g., \citealt{Dominik_2012, Bavera:2021, Zevin2021, Broekgaarden_2022, van_son_2022, Briel_2022_b}), this involves either limiting the mass accretion onto the primary black hole - during the phase of stable stellar mass transfer - to a sub-Eddington rate  (e.g., \citealt{van_son_2020, Zevin_Bavera_2022}) or fine-tuning certain free parameters during the common envelope evolution (CEE) phase of stellar evolution which lacks a full understanding (e.g., \citealt{Gallegos-Garcia:2021, Klencki:2021, Marchant:2021}).
In the case of stable mass transfer between binaries, limiting accretion to a sub-Eddington value causes the mass transfer to (at times) become non-conservative. This results in mass and angular momentum loss from the binary and thus a gradual hardening of the orbit. However, it is not clear if such an assumption is justified, with studies suggesting that super-Eddington accretion on black holes could be readily achievable (e.g., \citealt{2016_Sadowski, Ghodla:2023}).
On the other hand, CEE is a highly nonlinear process, and to attain a better understanding, 3D numerical simulation might be necessary. Meanwhile, conducting comprehensive multidimensional magnetohydrodynamic simulations of the CEE phase across a diverse range of spatiotemporal parameter spaces remains computationally too expensive (e.g., \citealt{Ricker:2008, Ricker:2012, Chamandy:2018, Moreno:2022, Lau:2022}). As such, presently, one has to resort to 1D studies, which, at times, may make ad hoc choices for the associated free parameters to produce BBHs with favorable orbital separation (e.g., \citealt{Breivik:2020, COMPAS_2022}). 

It is also disputed whether CEE could readily occur in close enough stellar binaries. In particular, in the \textit{detailed} 1D treatment of binary interaction, mass transfer can remain stable (e.g., 
 see \citealt{Gallegos-Garcia:2021, Klencki:2021, Marchant:2021} for analyses using \textsc{Mesa} models and \citealt{Briel_2022_b} using \textsc{Bpass}), which reduces the scope of CEE. As the latter plays a key role in hardening the orbit of the progenitors of BBHs, one can expect a reduced BBH merger rate estimate. Thus the question arises:  what other potential mechanism could help to harden a binary orbit? Three-body interactions (e.g., \citealt{Ziosi:2014, Dorozsmai:2024}) or the efficient inspiral of compact remnants in the disks of active galactic nuclei (e.g., \citealt{Leigh:2018, Kaaz:2023}) may offer such alternatives. 
 Separately, gravitational wave mergers may also occur via the dynamical capture of black holes in dense astrophysical environments (e.g., \citealt{Kulkarni_1993, Liu:2020, Gamba:2023}), which may not require an efficient hardening mechanism.

Since BBHs reside in a sea of interstellar medium (ISM), here we explore the effect of this surrounding baryonic medium on the orbital dynamics and merger properties of BBHs. Presently, most population studies assume that BBHs (born from twin stars) live in isolation (i.e., vacuum).  While this assumption is justified for a short evolutionary timescale, for longer timescales of $\mathcal{O}(10^9)$ yr (typical for LVK BBHs), this may strictly not be true. This is because, over such a large timescale, the minuscule impact of ISM interaction has a sufficient time to accumulate. As such, this work aims to quantify the characteristics of the baryonic medium surrounding the BBHs, under which it could play a important role in driving the LVK BBH mergers.

The ISM typically is an extremely tenuous fluid, containing nonrelativistic (charged and neutral) matter, relativistic charged particles known as cosmic rays, and magnetic fields (e.g., see \citealt{Ferriere:2001, Draine:2011} for a review).  
% 
% HI is abundant in galaxies typically traced by HI measured in absorption lines. The HI in WNM by the 21cm emission lines.
% 
A large fraction of the ISM in disk galaxies is in the form of neutral atomic hydrogen (HI) that can be broken into two parts: the cold neutral medium (CNM) with an average number density of $n_H$ = 20-50 cm$^{-3}$ and temperature $T \approx 100$~K and the warm natural medium (WNM) with $n_H = 0.6$ cm$^{-3}$ and $T = 5000$~K (e.g. \citealt{Draine:2011}). 
Under optimal circumstances, these two phases can coexist in pressure equilibrium, resulting in a two-phase medium (\citealt{Field:1969, Wolfire:1995, Wolfire:2003}, also see \citealt{McKee_Ostriker:1977}) containing relatively concentrated regions of CNM embedded in a diffused WNM.

Given that the progenitors of stellar black holes form in dense pockets of the H$_2$ regions, which themselves are the result of atomic-to-molecular gas conversion, the surroundings of these progenitor stars would presumably be rich in neutral hydrogen. However, because of the intense radiation emitted by the black hole forming OB-type star, much of the ISM surrounding the BBHs immediately after formation (including the circumstellar medium created by the progenitor's stellar winds and mass ejection during mass transfer) should be composed of warm neutral and warm ionized hydrogen (the latter also known as a warm ionized medium, WIM). 
The abundance of the latter type will depend on the amount of ionizing photons emitted by the progenitor stars and the strength of the supernova explosion (if any).

There is also the possibility that the supernova explosion expels a substantial portion of the ISM to large distances, thus creating a scarcity of interstellar matter around the subsequently formed black holes. However, the absence of core-collapse supernova observations from stars with mass $\gtrsim 18 M_{\odot}$ suggests that most black holes might form from a failed supernova explosion \citep{Smartt:2015}. When a successful explosion does occur, a rough estimate (using energy conservation) shows that matter displaced to a radial distance $d$ can fall back on the BBH in time 
\begin{equation}
    t \approx \frac{2 d^{3/2}}{3 \sqrt{GM}}  = 9.9 \left(\frac{d^3}{{\rm pc}}\right)^{3/2} \left(\frac{M}{M_\odot}\right)^{-1/2}  {\rm Myr} \,,
\end{equation}
where $M$ is the total mass of the BBH. For $M = 10 M_\odot$ and $d = 1$ pc, this gives $t \approx 3$ Myr, which~is negligible compared to the lifetime of many potential LVK BBHs, i.e., $\mathcal{O}(10^9)$~yr.

Furthermore, assuming no significant reheating of the surrounding ISM occurs post BBH formation, the time taken by the WNM to radiatively cool down can be estimated as
\begin{equation}
  t_{\rm cool} \approx \frac{n_H E}{\dot{\mathcal{C}}} \,,
\end{equation}
where $E = 3k_B T/2$ is its thermal energy, $T$ being the gas temperature and $\mathcal{\dot{C}}$ its mean cooling rate. The latter is defined as (e.g., \citealt{Draine:2011})
\begin{equation}
    \dot{C} = n^2_H \Lambda(T) \,,
\end{equation}
where $\Lambda$ is the cooling function that depends on the gas temperature and metallicity. For a medium constituting of HI gas, we assume \citep{Wolfire:2003}
\begin{equation}
    \Lambda(T) = 5.7 \times 10^{-26} \left(\frac{T}{10^4 K} \right)^{0.8} \, {\rm erg \, cm}^3 \, {\rm s}^{-1} \,.
\end{equation}
%
% \citep{Wadekar:2021}.
For typical values for the number density and temperature of the WNM (n$_H$ = 0.6 cm$^{-3}, T = 5000$ K), this yields $t_{\rm cool} \approx 2$  Myr (although this could be much higher in strongly metal deficient environments where cooling would be driven by Lyman-$\alpha$ transition which does not depend on the metal abundance, e.g., chapter 30 in \citealt{Draine:2011}). A similar time frame also exists for the cooling of WIM (e.g., see \citealt{Draine:2011}). Since we are interested in BBHs that live for $\mathcal{O}(10^9)$ yr, as such in this paper, we assume that, for the most part, the BBH surroundings always have some presence of a generic cold ISM\footnote{Later in Section~\ref{sec: Impact on BBH mass and merger rates}, we show that only a subset of BBHs (i.e. $\mathcal{O}(1)\%$) need to satisfy this criterion to contribute significantly to the LVK rates, while a higher percentage would result in inconsistency with the preexisting LVK observations.}. We further assume that these BBHs have a negligible velocity w.r.t. their surrounding medium given their presumably common source of origin (the case for a high relative velocity is discussed later in Section~\ref{sec: discussion}). 
The validity of these assumptions is a major source of uncertainty in the present work, and therefore, the formalism employed should only be viewed as a toy model of the ISM around BBHs. 

Under these assumptions for the ISM, we show that the properties of the LVK's BBHs (including their mass and merger rates) residing in such a medium could be significantly affected. On the other hand, if the assumptions of our toy model are strongly violated, one may safely assume the ISM to have a negligible impact on the BBH dynamics.
We also show that the accretion of WNM is inefficient and typically does not affect the BBH dynamics. Hence, the accretion of CNM will remain the focus of this paper. In all this, we ignore the effect of dark matter interaction with the BBHs and leave this analysis for a future study.

The remainder of this text is organized as follows. In Section~\ref{sec: Accretion of ISM on black holes}, we discuss the effect of ISM accretion on the BBH mass and orbital period. In Section~\ref{sec: Orbital decay of the binary}, we quantify the rate of the binary's orbital decay due to gravitational radiation emission, dynamical friction, and ISM accretion on the BBH. In Section~\ref{sec: Binary merger time}, we then calculate the resulting BBH merger time incorporating all these means of orbital decay. In Section~\ref{sec: Impact on BBH mass and merger rates}, we conduct a population study that allows us to quantify the effect of ISM on a generic cosmological population of BBHs and compare the result to the LVK observations. We end with a brief discussion in Section~\ref{sec: discussion} followed by a conclusion in Section~\ref{sec: conclusion}.

\vspace{-10pt}
\section{Accretion of ISM on black holes} \label{sec: Accretion of ISM on black holes}

\subsection{Effect on a black hole}

Consider a Schwarzschild black hole immersed in an isotropic surrounding of ISM. Such a fluid can be approximated with an ideal gas equation of state with pressure $P$ and density $\rho$ obeying the polytropic relation
\begin{equation}
    P = \frac{k_B T}{m_H} \rho^\gamma; \quad \gamma = 1 \,.
\end{equation}
Due to its gravitational influence the black hole would slowly accrete mass from the environment. The resulting mass accretion rate can be modeled using the Bondi-Hoyle accretion formula as \citep{Bond_Hoyle_1944}
\begin{equation}
    \dot{m}=\frac{4 \pi \lambda G^2 m^2 \rho_{\infty}} {\left(c_{\infty}^2+v_{\infty}^2\right)^{3 / 2}} \,,
    \label{eq: Bondi Hoyle}
\end{equation}
where $\rho_\infty$ is the density, $c_\infty$ is the speed of the sound, and $v_\infty$ is the velocity of the black hole w.r.t. the ISM. The subscript ``$\infty$'' indicates that the parameters are locally measured by an observer who is asymptotically far away from the black hole. This for us is the distance where the black hole's influence on the ISM becomes negligible. Additionally, the numerical factor \citep{Bondi_1952}
\begin{equation}
    \lambda = \frac{1}{4} \left( \frac{2}{5 - 3\gamma} \right)^{\frac{5 - 3\gamma}{2(\gamma - 1)}}; \quad \lambda(1) = e^{3/2} / 4 \approx 1.12 
\end{equation}
and 
\begin{equation}
    c_\infty = \sqrt{\frac{\gamma P_\infty}{\rho_\infty}} = \sqrt{\frac{k_B T_\infty}{m_H}} \,,
\end{equation}
where we have set $\gamma = 1$ in the final expression for $c_\infty$.

\subsubsection{Mass evolution of the black hole}

\begin{figure}
    \centering
    \vspace{-5pt}
    \includegraphics[width = 1\linewidth]{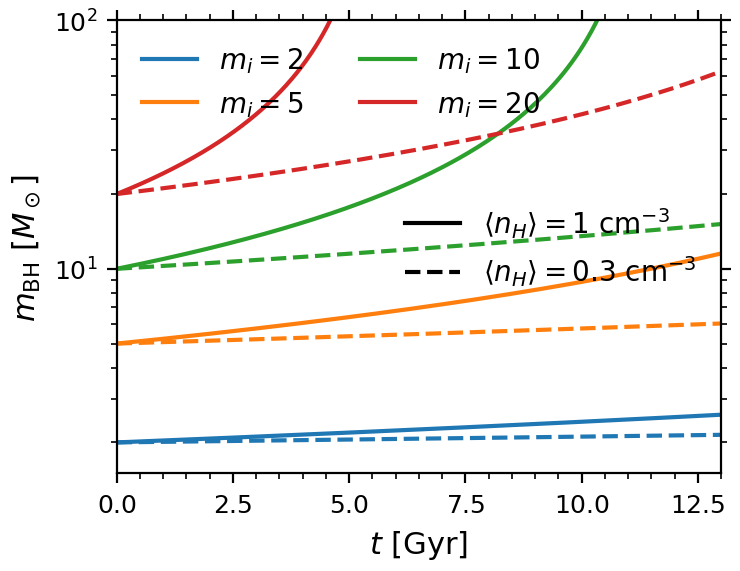}
    \vspace{-10pt}
    \caption{The mass evolution of a black hole embedded in a cold neutral medium of $T_\infty = 100$ K (see Eq.~\ref{eq: mass evolution}). The line styles represent the time-averaged particle number density of the CNM. The line color distinguishes black holes with different initial masses $m_i$. See Fig.~\ref{fig: mass evolution of a BH at larger T} for the case of a larger $T_\infty$. The case with $m_i = 2$ can be thought to instead represent a neutron star that accretes in the same fashion as a (horizon-possessing) black hole.}
    \label{fig: mass evolution of a BH}
\end{figure}

For a black hole comoving w.r.t. to its surrounding ISM, this implies $v_\infty = 0$ with the resulting expression in Eq.~\ref{eq: Bondi Hoyle} known as Bondi accretion rate \citep{Bondi_1952}. Solving Eq.~\ref{eq: Bondi Hoyle} for $v_\infty = 0$ thus gives us the mass evolution of the black hole as 
\begin{equation}
    m(t) =  \eta m_i \,; \quad \eta(t) = - \frac{1}{m_i \alpha t - m_i \alpha t_i -1}; \quad \eta(t) \geq 1 \,,
    \label{eq: mass evolution}
\end{equation}
% 
% \begin{equation}
%     \frac{-1}{m} + \frac{1}{m_i} =\frac{4 \pi G^2 \rho_\infty}{c_\infty^3} (t - t_i)\,,
%     \label{eq: mass evolution}
% \end{equation}
% 
where $m_i$ is the mass of the black hole at some initial time $t_i$ and $\alpha = \frac{4 \pi \lambda G^2 \rho_\infty}{c_\infty^3}$ (e.g., see Fig.~\ref{fig: mass evolution of a BH} for an illustration).

Eq.~\ref{eq: mass evolution} suggests that the mass $m$ of the black hole immersed in an ISM environment would approach infinity in a finite time $\Delta t:= t - t_i$. The latter can be calculated by setting the denominator in Eq.~\ref{eq: mass evolution} to zero, yielding
\begin{equation}
    \Delta t = \frac{c_\infty^3}{4 \pi \lambda G^2 \rho_\infty m_i} \,.
\end{equation}
Interestingly, even for a modest (time-averaged) value of $\langle n_H \rangle = 1$~cm$^{-3}$ and $T_\infty = 100$~K, this implies that the mass of a black hole with $m_i \gtrsim  8 \, M_\odot$ would \textit{diverge within a Hubble time }(e.g., Fig.~\ref{fig: mass evolution of a BH}).  On the other hand, if the surrounding constitutes a medium with a higher temperature, the black hole mass growth is strongly restricted, e.g. see Fig.~\ref{fig: mass evolution of a BH at larger T}. In particular, for $T_{\infty} \gtrsim 1000$ K, one can calculate that a black hole would experience negligible mass growth for the quoted values of $\langle n_H \rangle$. Thus, in this work, we focus only on a low-temperature environment.

\subsubsection{An upper bound on $\langle n_H \rangle$}

Even when the ISM has a low enough temperature, the supply of matter around the black hole would be limited, which would then set an upper limit on the mass of the black hole. Requiring that the latter remain finite allows us to set an upper bound on the value of $\langle n_H \rangle$. For example, if we are interested in time-averaging over a Hubble time, then setting $\Delta t = 1/H_0$ gives
\begin{equation}
    \langle n_H \rangle < \frac{H_0 (k_B T_\infty)^{3/2}}{4 \pi \lambda G^2 m_H^{5/2} m_i} \, \approx 8 \frac{ (T_\infty/100)^{3/2}}{m_i} \, {\rm cm}^{-3}\,,
    \label{eq: upper bound on n_H}
\end{equation}
where $m_i$ (in the RHS expression) is assumed to be in units of $M_\odot$. As expected, $\langle n_H \rangle$ is directly proportional to  $H_0, T_\infty$ and inversely proportional to $m_i$.

% Additionally, we have introduced the variable $n$ for later convenience in Section~\ref{sec: Impact on BBH mass and merger rates}. In the case when $\Delta t = 1$ and $T_\infty = 100$ K in Eq.~\ref{eq: upper bound on n_H}, $n$ can be viewed as mass-weighted particle number density in the unit time interval $\Delta t$.   

\subsection{Effect on a binary black hole} \label{sec: effect of ISM on BBH}

\subsubsection{Mass evolution of the BBH} \label{sec: Relative motion w.r.t. the ISM}

Now, let us consider a binary system composed of two Schwarzschild black holes embedded in an ISM. As earlier, we assume the binary as a whole to be coming w.r.t. to its surrounding ISM.
However, the binary would also rotate around its center of mass, thus giving rise to a non-zero $v_\infty$ for the individual black holes. 
Nevertheless, the gas is still asymptotically at rest w.r.t. the center of mass of the BBH. 

Let us introduce the length scale known as the transonic or Bondi radius 
\begin{equation}
    r_s = \frac{2GM}{c_\infty^2 }
\end{equation}
of the BBH configuration with total mass $M$. This corresponds to the radial distance from the center of mass of the BBH at which the accretion flow first becomes supersonic. Taking $r$ as the semi-major axis of the binary, if $r \ll r_s$ (which is always the case for LVK black holes\footnote{For example given a BBH with total mass $M = 10M_\odot$, and CNM with $T_\infty = 100$ K we get $r_s  \approx 4.6 \times 10^6 R_\odot$.}), then accretion at a large distance from the BBH can still be treated as Bondi accretion on a central compact object with mass $M$. 
On the other hand, the accretion dynamics near the binary could be complex. However, numerical simulations of binaries embedded in a gaseous environment with $r \ll r_s$ suggest that the accretion rate on the binary (as a whole) is \textit{enhanced} compared to accretion on two masses with the same total mass $M$ \citep{Farris:2010, Comerford:2019, Kaaz:2019}. In this work, we adopt a conservative approach, assuming each black hole in the binary accretes independently at the rate it would accrete in isolation as determined by setting $v_\infty = 0$ in Eq.~\ref{eq: Bondi Hoyle}.

\subsubsection{Impact on the BBH's orbital dynamics} 

The orbital angular momentum of an isolated binary can be expressed as
\begin{equation}
    L = \sqrt{G \mu^2 M r(1 - e^2)} \,,
    \label{eq: AM of binary}
\end{equation}
where $e, r$  and $\mu = m_1 m_2 / (m_1 + m_2)$ are the eccentricity, semi-major axis, and the reduced mass of the binary, respectively ($m_1, m_2$ being the component mass and $M = m_1 + m_2$).
If mass accretion remains isotropic - implying that the matter contains zero angular momentum at the radial asymptote, the inflowing material does not contribute to $L$. Therefore, $L$ should remain conserved (barring the loss in gravitational radiation). For simplicity, in this work, we set $e = 0$ which reduces $r$ to the separation between the black holes. This is a conservative estimate as a more eccentric orbit decays faster by the emission of gravitational radiation, leading to a more rapid inspiral.

To understand the orbital evolution of such a mass-accreting BBH, it would be sufficient to study how $r$ evolves with time.
Using Eq.~\ref{eq: mass evolution}, we can calculate the time-evolution of the expression $\mu^2 M$ as
\begin{align}
    \mu^2 M &= \left( \frac{\eta_1 \eta_2 m_{1,i} m_{2,i}}{\eta_1 m_{1,i} + \eta_2 m_{2,i}}\right )^2 ( \eta_1 m_{1,i} + \eta_2 m_{2,i}) \\
    & \geq \frac{(\eta_1 \eta_2 m_{1,i} m_{2,i})^2}{\eta_1 (m_{1,i} + m_{2,i})} = \eta_1 \eta_2^2 \mu_i^2 M_i \,,
    \label{eq: condition for AM conservation}
\end{align}
where in deriving the inequality, we have assumed $m_1 \geq m_2$ and hence $\eta_1 \geq \eta_2$.
Thus, from Eq.~\ref{eq: AM of binary} and \ref{eq: condition for AM conservation}, an estimate for the evolution of $r$ for an ISM accreting binary can be written as
\begin{equation}
    r \leq \frac{r_i}{ \eta_1 \eta_2^2} \,,
    \label{eq: separation evolution for fixed AM}
\end{equation}
where $r_i$ is the initial binary separation and the factor in the denominator is what causes $r$ to become time dependent. For the rest of this paper, we assume the equality in the above equation to hold. This implies a less rapid orbital decay.

We note that accretion of material can also occur from the circumstellar medium around the BBH, which is the leftover matter from the earlier episode of stellar evolution. In this case, the material might contain angular momentum, unlike the above-mentioned assumption. We discuss its impact on the BBH's orbital dynamics in Section~\ref{sec: discussion}.

\subsection{Feedback on the ISM} \label{sec: feedback from ISM}

The previous discussion highlights that the ISM temperature (at least for $r > r_s$) must remain low for it to significantly impact the evolution of BBHs (e.g., see Fig.~\ref{fig: mass evolution of a BH} and \ref{fig: mass evolution of a BH at larger T}). However, there exists a potential scenario where the energy released during the accretion process could heat up the surrounding ISM, thereby increasing its temperature.

\subsubsection{If the infalling matter does not form an accretion disk}

In the region with $r \leq r_s$, the isotropically infalling matter moves at a supersonic rate, thus preventing it from radiating energy through viscous dissipation. Therefore, this should not increase the temperature of the gas.
Meanwhile, the binary components with nonzero $v_\infty$ can stir up their surrounding gas due to their gravitational influence, creating local density variations around the orbit. This can induce turbulence, thus causing the gas to radiate and heat the surrounding ISM.
However, as mentioned previously, numerical studies of mass accretion in gravitationally bound systems \citep{Farris:2010, Comerford:2019, Kaaz:2019} suggest that this does not significantly affect the mass accretion rate. This may be attributed to the fact that in Eq.~\ref{eq: Bondi Hoyle}, the mass accretion rate on the BBH is {determined by the gas properties near} $r \gtrsim r_s$, which is much larger than the separation of the black holes. Once the gas falls past $r_s$, it is descending at a supersonic velocity (which further increases with infall). As a result, the gas should become more or less bound to the binary.

\subsubsection{If the infalling matter forms an accretion disk}

It is possible that the infalling gas borrows sufficient angular momentum from the BBH's orbit such that a fraction of it is assimilated into an accretion disk around the component masses. 
The characteristics of such an accretion disk will depend on the magnitude of the mass accretion rate on the individual black holes (e.g., \citealt{Frank_2002, Yuan_Narayan_2014}). For Bondi accretion on a black hole with mass $m$ (which holds at least at larger scales), the accretion rate scales as
\begin{equation}
    \dot{m} = 5.5 \times 10^{5} m^2 \cdot n_H \left( \frac{100}{T_\infty} \right)^{3/2} \,.
\end{equation}
Normalizing the above w.r.t. the Eddington accretion rate yields the fraction
\begin{equation}
   f = \epsilon \frac{\dot{m}}{ \Dot{m}_{\rm Edd}} = 3.9 \times 10^{-9}  \epsilon n_H  \left(\frac{m}{M_{\odot}}\right) \left( \frac{100}{T_\infty} \right)^{3/2} \,,
    \label{mass accretion w.r.t Eddintion rate}
\end{equation}
where $\epsilon$ is the accretion efficiency and $\dot m_{\rm Edd}$ is the Eddington accretion rate and takes the form
\begin{equation}
    \Dot{m}_{\rm Edd} = \frac{L_{\rm Edd}}{c^2}; \quad L_{\rm Edd} = \frac{4 \pi Gm c}{\kappa} \approx 1.26 \times 10^{38}\left(\frac{m}{M_{\odot}}\right) \mathrm{erg} \cdot \mathrm{s}^{-1} \,.
\end{equation}
Above, in calculating the last term, we assumed that the opacity $\kappa$ of the matter is purely driven by the accretion of ionized hydrogen.

Let us further assume that only a fraction $\nu$ of the accreted mass is assimilated into the accretion disk (with the rest getting accreted directly). Then, for the parameter values of $n_H = 1$ cm$^{-3}$ and $T_\infty = 100$ K, Eq.~\ref{mass accretion w.r.t Eddintion rate} suggest that the mass assimilation rate in the disk w.r.t. the Eddington rate should should evolve as 
\begin{equation}
   f = 3.9 \times 10^{-3} \epsilon \nu \left(\frac{m}{M_{\odot}}\right) \,.
\end{equation}
For illustrative purposes if we consider $\nu \in [0.1,1]$, $m \in [3 M_\odot, 40 M_\odot]$ and $\epsilon = 1/16$\footnote{As most stellar black holes might be slowly spinning at birth \citep{Fuller_Ma_2019}, we thus assume $\epsilon$ to be small.}, then one finds that for such a choice of $n_H$ and $T_\infty$, $f \in (6.3 \times 10^{-5}, 0.01)$.

Such values of $f$ span two solutions of the accretion flow, namely the radiatively inefficient accretion disks (e.g., \citealt{Narayan_Yi_1994, Blandford_1999, Yuan_Narayan_2014}) and the \cite{Shakura_Sunyaev_1973} accretion disks.
In the case of radiatively inefficient accretion flows, the temperature of matter is almost virial. The inflowing plasma develops a two-temperature state, where the ions contain most of the energy, but it is the electrons that are the more efficient cooling source. The ions and electrons do not efficiently thermalize; thus, the disk develops a two-temperature plasma. As the disk cannot cool, nearly all of the energy generated through viscous dissipation will be advected by the black hole \citep{Narayan_Yi_1994, Yuan_Narayan_2014} or a large fraction of the infalling matter will be lost in winds \citep{Blandford_1999, Yuan_Narayan_2014} (will part of the lost matter not significantly contributing towards $f$ either). 
Such an accretion flow may form a corona at the inner edge of the accretion disks and thus glow in hard X-rays. However, the absorption cross-section of (hard) X-rays for low-metallicity matter is relatively small \citep{Wilms:2000}. As such, they might not significantly heat the ISM surrounding the black holes, implying that a steady accretion rate could be maintained. Additionally, if the disk develops winds, the gas will also remove with it its acquired angular momentum, thus helping in hardening the binary without increasing the mass of the black hole.

In the scenario where the accretion flow becomes radiatively efficient (that is, $f \gtrsim 0.01$, e.g., \citealt{Yuan_Narayan_2014}), it will begin to radiate black-body-like radiation. For such a case, ISM heating could be significant and further mass will be prevented from efficiently assimilating in the disk. In such a scenario,  the accretion process may still proceed but with a duty cycle. Initially, mass is assimilated in the disk, which eventually reaches the black hole on a vicious timescale. This mass then heats up the surrounding ISM, thus halting further mass accretion. The ISM then gradually cools over a few million years, allowing for an accretion disk to form again, thus continuing the cycle. As this may reduce the mass accretion rate, thus the BBH's evolution may not be affected significantly. We note that large $f$ values are more likely for massive black holes. 

The efficiency at which the infalling matter becomes part of the accretion disk rather than getting directly accreted remains a major caveat in the current work. In the following, we assume that $\nu$ or $f$ is small, which then allows us to ignore the effect of ISM heating in the vicinity of the black holes. The validity of the results shown in the following should be interpreted within the domain of this assumption.

\section{Orbital decay of the binary} \label{sec: Orbital decay of the binary}

\subsection{Decay due to gravitational radiation with time-varying masses} \label{subsec: decay due to GW emission}

The orbital decay of a BBH due to the emission of gravitational radiation in the weak field regime is governed by the time variation of the quadrupole moment tensor
\begin{equation}
    Q_{i j}=  m_{1}x_{1, i} x_{1, j} + m_{2}x_{2, i} x_{2, j} \,,
    \label{eq: Quadrupole moment}
\end{equation}
where $x$ is the position of the point mass. The rate of radiative loss of binary's orbital energy can then be written as \citep{Peters_1964}
\begin{equation}
    \left.\frac{d E}{d t}\right|_{\rm GW}=\frac{G}{5 c^{5}}\left(\frac{d^{3} Q_{i j}}{d t^{3}} \frac{d^{3} Q_{i j}}{d t^{3}}-\frac{1}{3} \frac{d^{3} Q_{i i}}{d t^{3}} \frac{d^{3} Q_{j j}}{d t^{3}}\right) \,,
\end{equation}
where the subscript ``GW'' stands for gravitational wave. For a circular orbit, this reduces to 
\begin{equation}
    \left.\frac{d E}{d t}\right|_{\rm GW}=-\frac{32}{5} \frac{G^{4} \mu^{2} M^{3}}{c^{5} r^{5}}+\mathcal{O}(\dot{M}) \,.
    \label{eq: energy emission rate by GW}
\end{equation}
We note that due to time-varying masses, the first term on the RHS of the above equation \textit{is at least larger than} the one given in \cite{Peters_1964} by a factor of $\eta_1 \eta_2^{4}$ (cf. Eq.~\ref{eq: condition for AM conservation}) which is implicit in the mass term $M$. This results in a relatively larger loss of energy in gravitational radiation, which further increases monotonically with time.  Additionally, the last term also appears due to the time-varying mass of the black holes in Eq.~\ref{eq: Quadrupole moment}. As this term only has a sub-leading effect, it is thus ignored. If we now write the total orbital energy of the binary as  
\begin{equation}
    E = - \frac{G\mu M}{2r} \,,
    \label{eq: energy of binary}
\end{equation}
then Eq.~\ref{eq: energy emission rate by GW} and \ref{eq: energy of binary} together give us the orbital decay rate due to gravitational wave emission as
\begin{equation}
    \left.\frac{d r}{d t}\right|_{\rm GW}=\left.\frac{2 r^{2}}{G \mu M} \frac{d E}{d t}\right|_{\rm GW}=-\frac{64}{5} \frac{G^{3} \mu M^{2}}{c^{5} r^{3}}
\end{equation}
with $M$ now being time-dependent.

\subsection{Decay due to accretion of (asymptotically) zero angular momentum matter}

As the accreted material adds mass but not angular momentum to the BBH, the orbit of the binary would gradually decay. While it is possible that the motion of the BBH could potentially stir up the ISM (thus injecting angular momentum into the surrounding media), the source of this angular momentum ultimately originates from the BBH’s orbit. Eventually, the surrounding material will be accreted by the BBH. It is thus fair to assume that, outside of the angular momentum loss in gravitational radiation, the orbital angular momentum of the mass-accreting binary remains conserved. 

Consequently, we use Eq.~\ref{eq: separation evolution for fixed AM} to calculate the orbital decay rate due to isotropic mass accretion as
\begin{equation}
     \left.\frac{d r}{d t}\right|_{\rm AM} = -\left[ \frac{\dot{\eta}_1}{\eta_1}   + \frac{2 \dot{\eta}_2}{\eta_2}    \right] r \,,
     \label{eq: orbital decay due to AM conservation}
\end{equation}
where the subscript ``AM'' implies that the decay occurs due to the conservation of the BBH's angular momentum.

\begin{figure}
    \centering
    \vspace{-5pt}
    \includegraphics[width = 1\linewidth]{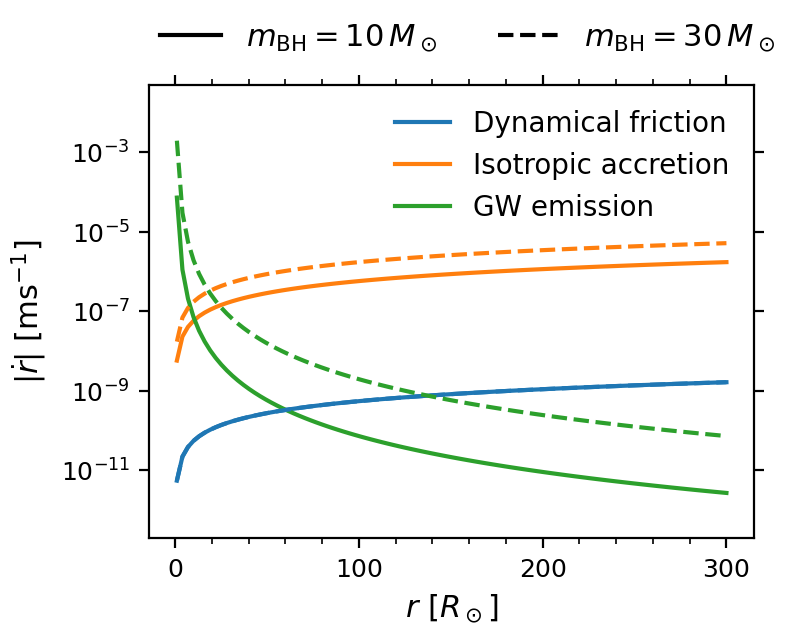}
    \vspace{-10pt}
    \caption{The absolute rate of orbital decay of equal component mass BBHs at different separations. The line styles represent the two masses considered. The line color represents the physical mechanism driving the orbital decay. Here, we consider a particle density of $\langle n_H \rangle = 1$ cm$^{-3}$. {The two line styles for the blue curve} (i.e., dynamical friction) apparently \textit{overlap} due to the low resolution on the y-axis. }
    \label{fig: rate of orbital decay}
\end{figure}

\subsection{Decay due to dynamical friction}

As discussed in Section~\ref{sec: Relative motion w.r.t. the ISM}, even a BBH comoving with the ISM will still have components with $v_\infty \neq 0$. Thus due to their interaction with the ISM, the black holes in the binary would consequently experience a gravitational drag force, also known as dynamical friction \citep{Chandrasekhar:1943}. Under the assumption that the surrounding medium constitutes a cold ISM with $T_\infty = 100$ K, all the black holes of interest would be moving at a supersonic speed w.r.t. the background medium. 

The force due to dynamical friction on a mass moving from a uniform background medium  can be approximated as \citep{Ostriker:1999}
\begin{equation}
    F_{\text {DF}}=\frac{4 \pi G^2 m^2 \rho_\infty  \mathcal{I}}{v_\infty^2} \,,
    \label{eq: force due to dynamical friction}
\end{equation}
where $\mathcal{I}=\ln \left(1-1 / \mathcal{M}^2\right) / 2+\ln \left(b_{\max } / b_{\min }\right)$. Here, $\mathcal{M} = v_\infty / c_\infty$ is the Mach number of the problem with $\mathcal{M} \gg 1$ for supersonic motion.

Additionally, $b_{\rm max}, b_{\rm min}$ correspond to the minimum and maximum impact parameters of the interaction. We choose $b_{\rm min}$ equal to the accretion radius of the black hole (also known as the Bondi-Hoyle-Lyttleton radius), i.e.,  $b_{\rm min} = 2Gm/v_\infty^2$, while $b_{\rm max}$ remains a free parameter in the present work and we set $b_{\rm max} = v_\infty \Delta t$, $\Delta t$ being the time elapsed since the black hole was born. This allows us to calculate the energy dissipation rate as 
\begin{equation}
     \left.\frac{d E}{d t}\right|_{\rm DF} = - v_\infty \cdot F_{\rm DF} = - \frac{4 \pi G^2 m^2 \rho_\infty \mathcal{I}}{v_\infty} \,.
\end{equation}
Using the above in conjunction with Eq.~\ref{eq: energy of binary} then yields
\begin{equation}
    \left.\frac{d r}{d t}\right|_{\rm DF} =  - \frac{8 \pi G^{1/2} m \rho_\infty \mathcal{I} r^{5/2}}{M^{3/2}} \,.
    \label{eq: decay rate due to dynamical friction}
\end{equation}

 In the case of a black hole in a binary, the immediate surroundings of the black hole will differ from that assumed in deriving Eq.~\ref{eq: force due to dynamical friction} (e.g., see Section~\ref{sec: feedback from ISM}). In particular, for such black holes, a large wake formation (behind the black holes) might be prevented which will lower the impact of dynamical friction on the black holes. As we discuss below, dynamical friction has only a sub-leading effect on the BBH dynamics, and thus, an exact treatment is not necessary.

\section{Binary merger time} \label{sec: Binary merger time}

Given the various forms of orbital decay discussed in the previous section, we now calculate the resulting BBH merger time.
By differentiating Eq.~\ref{eq: energy of binary} w.r.t. time, we find the total orbital decay rate as
\begin{equation}
    \frac{d r}{d t}=\left.\frac{d r}{d t}\right|_{\rm GW}+\left.\frac{d r}{d t}\right|_{\rm AM} + \left.\frac{d r}{d t}\right|_{\rm DF}
    \label{eq: total decay rate}
\end{equation}
The individual components of the above equation are plotted in Fig.~\ref{fig: rate of orbital decay}. In particular, it shows that at \textit{all binary separations}, the decay term due to dynamical friction is dominated by at least one of the other two mechanisms\footnote{Fig.~\ref{fig: rate of orbital decay} only considers a subset of black hole masses and a fixed $\langle n_H \rangle$. However, one can check that the decay due to dynamical friction should remain subdominant for other relevant values of these parameters. Also, we note that in calculating the decay due to dynamical friction, we have generously set $\Delta t = 10^9$ yr in the expression for $b_{\rm max}$ in Eq.~\ref{eq: decay rate due to dynamical friction}.}. 
As such, for ease of calculation, we safely ignore the decay resulting from dynamical friction. Moreover, qualitatively, one may assume that over long timescales, for BBH that comoves with the ISM, any momentum imparted to the gas by the BBH will eventually be accreted by them, hence negating the effect of dynamical friction.
Defining 
\begin{equation}
    \beta_{i}=\frac{64}{5} \frac{G^{3} \mu_{i} M_{i}^{2}}{c^{5}} \,,
\end{equation}
where the subscript $i$ indicate value at $t_i$ then Eq.~\ref{eq: total decay rate} yields
\begin{equation}
    \frac{d r}{d t} = -\frac{\beta_{i}}{r^{3}} \eta_1 \eta_2^2 -\left[ \frac{\dot{\eta}_1}{\eta_1}   + \frac{2 \dot{\eta}_2}{\eta_2}    \right] r \,.
\end{equation}
This can be rewritten as 
\begin{equation}
    \frac{d r^4}{d t}  + \underbrace{4 \left[ \frac{\dot{\eta}_1}{\eta_1}   + \frac{2 \dot{\eta}_2}{\eta_2}    \right] }_{P(t)} r^4 = \underbrace{- 4 \beta_i \eta_1 \eta_2^2 }_{Q(t)} \,.
    \label{eq: linear diff eqn}
\end{equation}
This is a first-order linear ordinary differential equation, which on substituting $y(t) = r^4(t)$, takes the form
\begin{equation}
    \dot{y}(t)+P(t) y(t)=Q(t) \,.
\end{equation}
This equation has the exact solution
\begin{equation}
    y(t)=e^{-\int_{t_{i}}^{t} P\left(t^{\prime}\right) d t^{\prime}}\left[\int_{t_{i}}^{t} e^{\int_{t_{i}}^{t^{\prime}} P\left(t^{\prime \prime}\right) d t^{\prime \prime}} Q\left(t^{\prime}\right) d t^{\prime}+y\left(t_{i}\right)\right] \,,
\end{equation}
where the integration factor takes the form
% 
% \begin{equation}
%     e^{ \pm \int_{t_i}^{t} P\left(t^{\prime}\right) d t^{\prime}} = \left[ \frac{\eta_1(t)}{\eta_2(t_i)}\right]^{\pm 4} \left[ \frac{\eta_2(t)}{\eta_2(t_i)}\right]^{\pm 8} \,,
% \end{equation}
% 
\begin{equation}
    e^{ \pm \int_{t_i}^{t} P\left(t^{\prime}\right) d t'} = \eta_1^{\pm 4}(t) \eta_2^{\pm 8}(t) \,.
\end{equation}
% 
% Above, without loss of generality, we have chosen $t_i = 0$ and by definition $\eta(t_i) = 1$.
Above, by definition, we have set $\eta(t_i) = 1$, yielding
\begin{equation}
    y(t) = \eta_1^{- 4}(t) \eta_2^{- 8}(t)  \left[y(t_i) - 4 \beta_i \int_{t_i}^{t} \eta_1^5(t') \eta_2^{10}(t') dt' \right] \,.
    \label{eq: Linear ODE}
\end{equation}
The merger time of an ISM accreting BBH is then the timescale $T_{\rm tot}$ (the subscript implies net decay due to the various mechanisms detailed in Eq.~\ref{eq: total decay rate}) such that 
\begin{equation}
    \lim _{t \to T_{\rm tot}} y(t) = 0 \,,
    \label{eq: condition for merger}
\end{equation} 
while the merger time of a BBH \textit{living in vacuum} - i.e., purely due to emission of gravitational radiation - is \citep{Peters_1964}
\begin{equation}
    T_{\rm GW} = \frac{y(t_i)}{4 \beta_i}  = \frac{5c^5 r_i^4}{256 G^3 \mu_i M_i^2}\,.
    \label{eq: Peters merger time}
\end{equation}
Combining Eq.~\ref{eq: Linear ODE} and \ref{eq: condition for merger} then gives the relation between $T_{\rm tot}$ and $T_{\rm GW}$ as
\begin{equation}
    \int_{t_i}^{T_{\rm tot} + t_i}  \eta_1^5(t) \eta_2^{10}(t) dt = T_{\rm GW} \,.
    \label{eq: general merger time}
\end{equation}

\begin{figure}
    \centering
    \vspace{-5pt}
    \includegraphics[width = 1\linewidth]{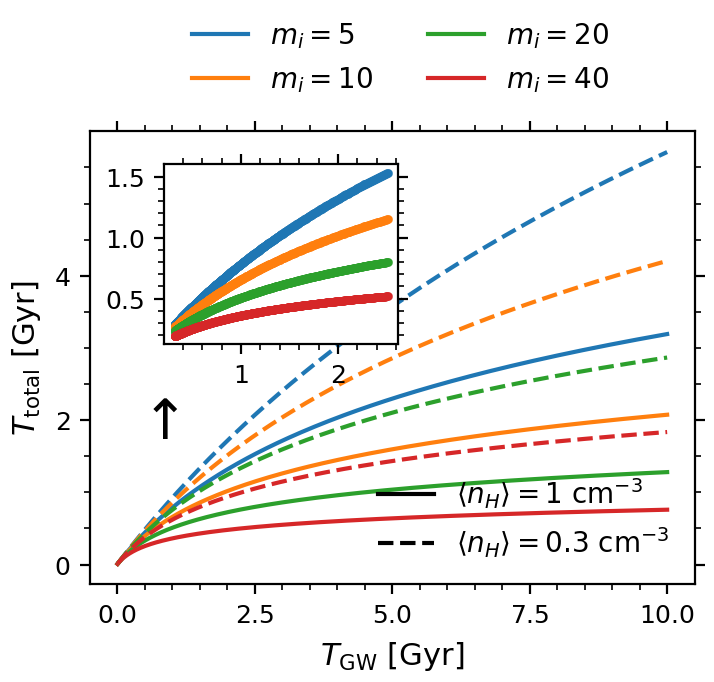}
    \vspace{-10pt}
    \caption{The merger time of equal component mass ISM accreting BBHs as a function of their (purely) gravitational wave emission induced merger time (see Eq.~\ref{eq: merger time}). The colors represent the different initial black hole masses in the binary (with mass-ratio = 1). The line styles represent the time-averaged particle density of the surrounding ISM. The ISM temperature is set to $T_\infty=100$ K.}
    \label{fig: Merger time}
\end{figure}

The above integral can be evaluated analytically using symbolic algebra tools but is not presented here due to its lengthy size\footnote{We note that the hence acquired solution starts to diverge as the mass ratio of the binary $q:= m_2/m_1 \gtrsim 0.85$. Thus, in Section~\ref{sec: Impact on BBH mass and merger rates}, we interpolate between this general solution and the solution presented in Eq.~\ref{eq: merger time} for $q \in (0.8,1]$. The general solution can be found in the jupyter notebook URL included in the Data Availability Statement.}. For comparative purposes, we instead provide the solution for the special case when $m_{1,i} = m_{2,i} = M_i/2$ (and set $t_i = 0$) yielding
% 
% 
% \begin{equation}
%    T =  \frac{1}{m_i \alpha} \left[-   \left(\frac{1}{14 m_i \alpha T_{\rm GW} + 1} \right)^{1/ {14}} + 1\right] \,.
% \end{equation}
% 
\begin{equation}
    T_{\rm tot} = \frac{1}{m_i \alpha} \left[1 - \left(\frac{1}{14m_i \alpha  T_{GW}+ 1}\right)^{1/{14}} \right] \,.
    \label{eq: merger time}
\end{equation}
Since $T_{\rm GW}$ can be readily calculated for a binary in the weak field limit, thus knowing the values of the initial mass of the black hole $m_i$, the ISM temperature at large distance from the BBH $T_\infty$ and time-averaged particle density $\langle n_H \rangle$, Eq.~\ref{eq: merger time} makes the calculation of $T_{\rm tot}$ straightforward. 
The solution to Eq.~\ref{eq: merger time} is illustrated in Fig.~\ref{fig: Merger time}, showing the quantitative effect of an ISM environment on the BBH merger time.

\section{Impact on LVK Observables} \label{sec: Impact on BBH mass and merger rates}

To estimate the impact of the ISM environment on the BBH merger properties, we apply the previous calculation for mass growth (Eq.~\ref{eq: mass evolution}) and accelerated orbital decay (Eq.~\ref{eq: general merger time}) to a realistic population of stellar black holes. For the latter, we use the \textsc{Bpass} v2.2.1 detailed binary stellar models \citep{Bpass2017, Bpass_2018}, which are then distributed over a range of formation redshifts. This distribution is a function of the redshift-dependent star formation rate (weighted with an initial mass function) and metallicity evolution of the Universe. Over the course of the subsequent stellar evolution, episodes of mass transfer and CEE may cause the orbit of these binaries to harden, resulting in BBH systems that could be potential sources for the LVK detectors \citep{LIGO_instrument_paper, Virgo_instrument_paper, KAGRA_instrument_paper}. 

For more details on the implemented {stellar} population synthesis, we refer the reader to Appendix A in \cite{Ghodla_coupled_black_holes:2023} while the BBH merger rate calculation is discussed in Appendix \ref{sec: rate calculation} of the current work.

\subsection{Stable mass transfer and common envelope evolution in stellar binaries}

In separate studies \citep{Briel_2022_a, Briel_2022_b, Ghodla_coupled_black_holes:2023}, it has been observed that \textsc{Bpass} models underpredict the BBH merger rate. This can be attributed to the treatment of mass transfer episodes between progenitors of BBHs. Firstly, in the detailed treatment of the \textsc{Bpass} models, mass transfer is relatively more stable than in other rapid population synthesis codes that treat the interaction phase between the stellar binaries parametrically (e.g., \citealt{Breivik:2020, COMPAS_2022}). However, this is not just limited to \textsc{Bpass}, as similar effect has also been observed in the detailed models of \textsc{Mesa} \citep{Gallegos-Garcia:2021, Klencki:2021, Marchant:2021}. Additionally, in \textsc{Bpass} super-Eddington accretion onto the black hole (from the secondary stellar companion) is allowed. As such, when the mass transfer is stable, it also largely remains conservative, which reduces the scope for orbital hardening \citep{Briel_2022_b}. Secondly, the \textsc{Bpass} models undergo more efficient\footnote{By efficient, we mean the binary efficiently loses the envelope during CEE.} CEE episodes as compared to the rapid population study codes. Since CEE plays a key role in hardening the orbit of the progenitors of BBHs, one can expect a reduced BBH merger rate estimate.
% 
% 
% Moreover, in less efficient treatment of CEE as at times employed in the rapid population synthesis codes that treat the interaction phase between the stellar binaries parametrically (e.g., \citealt{Breivik:2020, COMPAS_2022}), uncertainty remains regarding the choice of available free parameters, resulting in a range of BBH merger rates (e.g., \citealt{Broekgaarden_2021, Broekgaarden_2022}).

For the current work, we take the \textsc{Bpass} models as a template and consider the impact of an ISM environment on its BBHs.

\subsection{Redshift evolution of $\langle n_H \rangle$}

As the galaxies in the early universe were presumably more gas-rich, thus the supply of gas surrounding the BBHs is expected to last longer for those formed at higher redshifts $z$. This implies that $\langle {n_H} \rangle$ should be a monotonically increasing function of $z$. Here, we assume this to take the form of a power-law as
\begin{equation}
    \langle n_H \rangle(z)  =  \langle {n_H} \rangle_0 (1 + z)^\kappa \,,
    \label{eq: particle number density with redshift}
\end{equation}
where $z$ represents the redshift at BBH formation. The term $\langle {n_H} \rangle_0$ can be seen as $\langle {n_H} \rangle$ evaluated at $z = 0$ or equivalently as $\langle {n_H} \rangle$ when there is no redshift dependence. 

In the simulation result shown in the following sections, we adopt two values for $\kappa$, namely $\kappa = 1$ and~3 and set $T_\infty = 100$ K assuming a CNM surrounding.  Moreover, only some BBHs may be found in an ISM environment that matches the requirements of our toy model. Thus, we present results assuming that of all BBHs formed in our simulation, only $\mathcal{O}(1)$\% of them are embedded in a favorable ISM surrounding while the rest either live in a vacuum or do not match the requirement of our toy model to experience a noticeable impact on their dynamics. The choice of this subset of $\mathcal{O}(1)$\% BBHs in the simulation is made on a random basis and contains no redshift bias.

Finally, although we do not consider the following scenario here, however instead of assuming that $\langle n_H \rangle_0$ takes the same value for all BBHs born in the simulation irrespective of their initial mass $M_i$, one may assume that due to a limited supply of matter, the surroundings of more massive black holes become ISM-poor at a faster pace (as massive binaries accrete at a faster rate, see Eq.~\ref{eq: Bondi Hoyle} and Fig.~\ref{fig: mass evolution of a BH}) and thus will have a lower value of $\langle n_H \rangle_0$. Additionally, for more massive black holes, a larger $ \langle n_H \rangle_0$ also would raise the value of $f$ at a relatively faster rate, thus causing ISM heating that will limit the accretion rate (see Section~\ref{sec: feedback from ISM}). Thus, to keep $n_H$ low, one could employ a mass weighting in Eq.~\ref{eq: particle number density with redshift} and reformulate it as (cf. Eq.~\ref{eq: upper bound on n_H})
\begin{equation}
    \langle n_H \rangle =  n \frac{ (T_\infty/100)^{3/2}}{M_i} (1+z)^\kappa \, {\rm cm}^{-3}\,,
    \label{eq: mass weighting}
\end{equation}
where $n$ is now the BBH mass-corrected variant of $\langle n_H \rangle_0$ that determines the time-averaged particle number density.

\begin{figure}
    \centering
    \vspace{-5pt}
    \includegraphics[width = 1\linewidth]{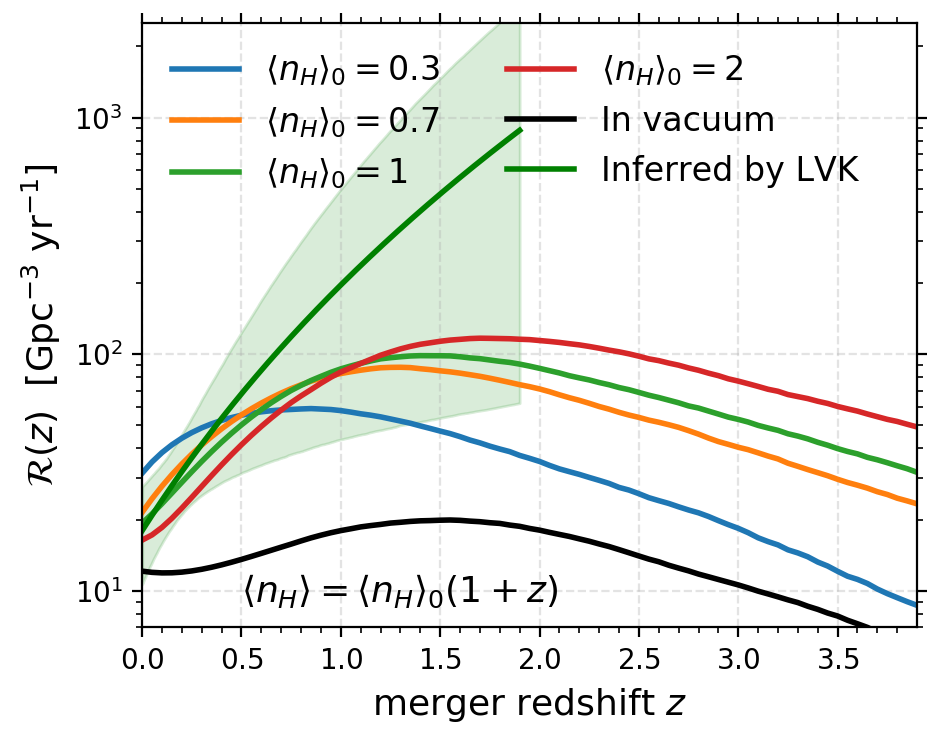}
    \includegraphics[width = 1\linewidth]{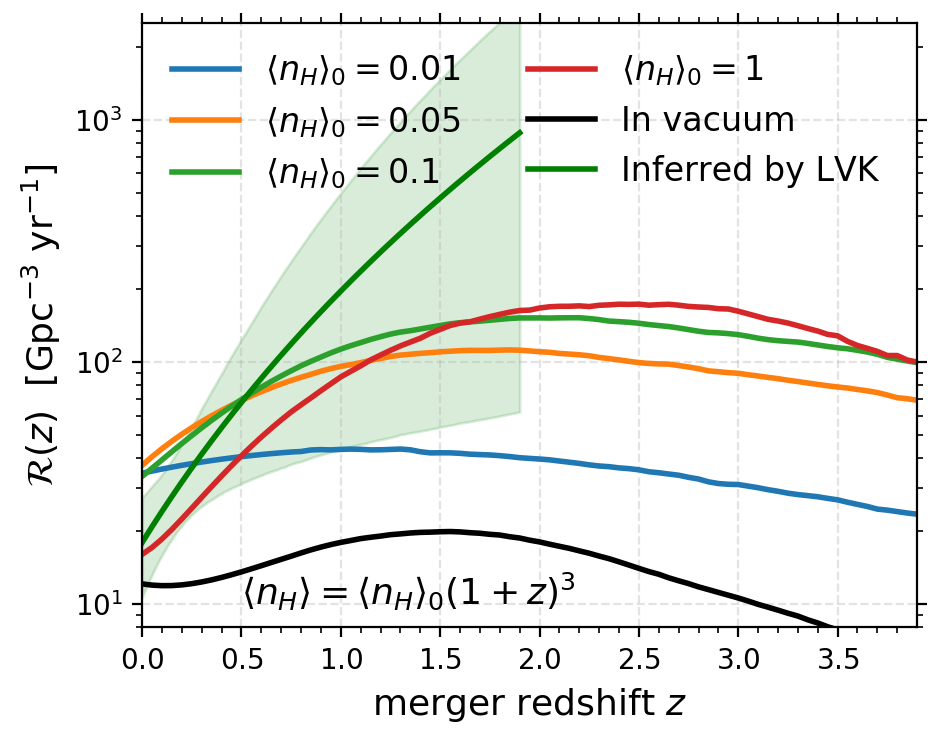}
    \vspace{-10pt}
    \caption{The source frame merger rate for the various employed $\langle n_H \rangle_0$ values. The top figure is for a linear redshift dependence (see Eq.~\ref{eq: particle number density with redshift}), while the bottom is for a cubic dependence. The top figure assumes only 1.5\% of the BBHs have an optimal ISM surrounding, while the bottom assumes this to be 3\%. These choices have been made to yield merger rates within the green band. The green curve shows the expected BBH merger rate as inferred by LVK, while the shaded region shows its 90\% credible interval (data taken from \citealt{GWTC_3_data_set}). The black curve is the BBH merger rate calculated assuming all BBHs live in a vacuum.}
    
    \label{fig: volumetric merger rate}
\end{figure}

\vspace{-5pt}
\subsection{Source-frame BBH merger rate}

More details on the source frame BBH merger rate calculation are presented in Appendix \ref{subsec: source frame rate} and \ref{subsec: A2}, while the resulting rates are shown in Fig.~\ref{fig: volumetric merger rate}. Here, we assume that only 1.5\% (top figure) and 3\% (bottom figure) of the BBHs reside in an optimal ISM environment for their properties to be affected. These choices have been made to yield merger rates within the green band shown in the figure. 
A higher percentage would result in a merger rate that starts to diverge from the LVK inferred range\footnote{We find that on changing this percentage, the shape of the curve remains preserved, but with the merger rate is vertically displaced. This holds as long as the percentage is not dependent on the formation redshift of the BBHs.}. This implies that only a small fraction of the BBHs within our simulation should satisfy the condition of a favorable ISM environment. 
Separately, differences in implemented physics, such as the treatment of CEE or mass transfer stability, can affect the number of BBHs that can survive in the \textsc{Bpass} code compared to other population synthesis models. As more binaries would lead to more mergers, thus a variation in stellar evolution physics can also affect the choice of the aforementioned percentage between population studies.

Comparing the results presented in Fig.~\ref{fig: volumetric merger rate} to the case where BBHs evolve in a vacuum (black curve) shows the effectiveness of an ISM surrounding in hardening the BBH orbit. 
At the same time, the merger rate curve also approximately follows the expected trend. The shape of the curve and the strength of the merger rate are dependent on the value of $\langle n_H \rangle$. Depending on the redshift evolution of the latter quantity, we find that an $\langle n_H \rangle_0 \approx 0.1-2$ cm$^{-3}$ results in a fair approximation to the LVK inferred rate when $\kappa = 1$. However, a smaller $\langle n_H \rangle_0$, e.g., $\langle n_H \rangle_0 \lesssim 0.1$ leads to deviation from the LVK inferred trend. Similarly for $\kappa = 3$, $\langle n_H \rangle_0 \approx 0.01 - 1$ cm$^{-3}$ also results in a fair approximation to the LVK inferred rate with $\langle n_H \rangle_0 \lesssim 0.01$ resulting in deviation. On the other hand, large values of $\langle n_H \rangle_0 $ (including some shown in Fig.~\ref{fig: volumetric merger rate}) may be disfavored on physical grounds as that would require a strong and continuous supply of ISM and such $\langle n_H \rangle_0 $ could lead to significant ISM heating (see Section~\ref{sec: feedback from ISM}, especially for massive black holes.

\subsection{Detectable BBH mergers}

The detectable number of BBH mergers as a function of their source frame mass is shown in Fig.~\ref{fig: detectable merger rate} (for more details on the employed methodology, see Appendix \ref{subsec: intrinsic rate}). To assist in readability, only a subset of the $\langle n_H \rangle_0$ values shown in Fig.~\ref{fig: volumetric merger rate} are plotted in Fig.~\ref{fig: detectable merger rate}. The figure shows the number of detections that one may expect to observe on Earth in a year, given all three LVK detectors work in quadrature for a full year at the third observing run (O3) sensitivity with the noise curves taken from \cite{LVK}. We note that in reality, the KAGRA detector only became operational in the last phase of the third observation run.

We find that there is a tendency for the ISM-accreting BBHs to be relatively more massive than the preexisting observations. One way to prevent an excess of such systems could be by assuming a lower value of $\langle n_H \rangle_0$ for more massive black holes - or employing the mass-weighting approach presented in Eq.~\ref{eq: mass weighting}. This is because, due to a limited supply of matter, the surroundings of more massive black holes become ISM-poor at a faster pace (as massive binaries accrete at a faster rate, see Eq.~\ref{eq: Bondi Hoyle}). Moreover, for large $\langle n_H \rangle_0$ values, more massive black holes would generate stronger radiation feedback on their surrounding ISM. Thus, this should result in smaller $\langle n_H \rangle_0$ values being more favorable, which in turn will lower the mass of such black holes.

A large fraction of these ISM-accreting BBHs are sufficiently massive that one or both of their components fall within the pair-instability mass-gap \citep{Fowler_and_Hoyle1964}. This mass-gap arises when massive stars develop the right conditions for efficient $e^-e^+$ pair-production, which leads to a complete disruption of the star. As such, no black hole remnant would be left behind.
Finally, if the ISM-accreting black holes do not gain significant angular momentum, they would develop relatively small values of the dimensionless spin parameter as compared to black holes living in a vacuum. This is in line with a majority of LVK observations to date \citep{Abbott_O3a_2021, Abbott_O3b_2021}. However, this might not remain true if the infalling gas acquires angular momentum for the BBH orbital dynamics, which is subsequently accreted by the black holes. There are also suggestions that while the matter gets accreted from the disk, the presence of large-scale magnetic fields near the horizon of the black hole could result in the formation of bipolar jets. The jets can feed on the spin angular momentum of the black hole, thus resulting in a spin down (e.g., \citealt{Tchekhovskoy_2011}).

\vspace{-5pt}
\section{Discussion} \label{sec: discussion}

\subsection{What if the BBHs have a nonzero $v_\infty$?}

Till now, we have assumed that, for the most part, the BBHs are at rest w.r.t. their surrounding ISM. Here, we qualitatively discuss the impact of a nonzero $v_\infty$ on our analysis.

\subsubsection{Impact on mass} \label{subsec: Impact of mass of a non-zero velocity}

From Eq.~\ref{eq: Bondi Hoyle} it can be seen that a BBH moving with a nonzero $v_\infty$ w.r.t. the surrounding medium would experience a reduction in the mass accretion rate. If we assume that $v_\infty = k c_\infty$, where $k \in \mathbb{R}$, then it follows that 
\begin{equation}
    \dot{m}_{v_\infty \neq 0} =  \dot{m}_{v_\infty = 0} \frac{1}{(k^2 + 1)^{3/2}} \,.
    \label{eq: mass accretion rate reduction}
\end{equation}
In the case where the motion is subsonic (i.e., $k < 1$), Eq.~\ref{eq: mass accretion rate reduction} suggests an utmost reduction in $\dot{m}$ by a factor of $\approx 2.8$. As such, there is a fair possibility that the masses of subsonically moving BBHs could be influenced by the ISM\footnote{For a CNM with $T_\infty=100$ K, $c_\infty \approx 0.9$ kms$^{-1}$.}. On the other hand, even for a mild supersonic velocity of $v_\infty = 2 c_\infty$, $\dot{m}$ falls by a factor of $\approx 11$. As such, BBHs moving at relatively higher velocities experience a diminishing impact from the ISM.

\begin{figure}
    \centering
    \vspace{-5pt}
    \includegraphics[width = 1\linewidth]{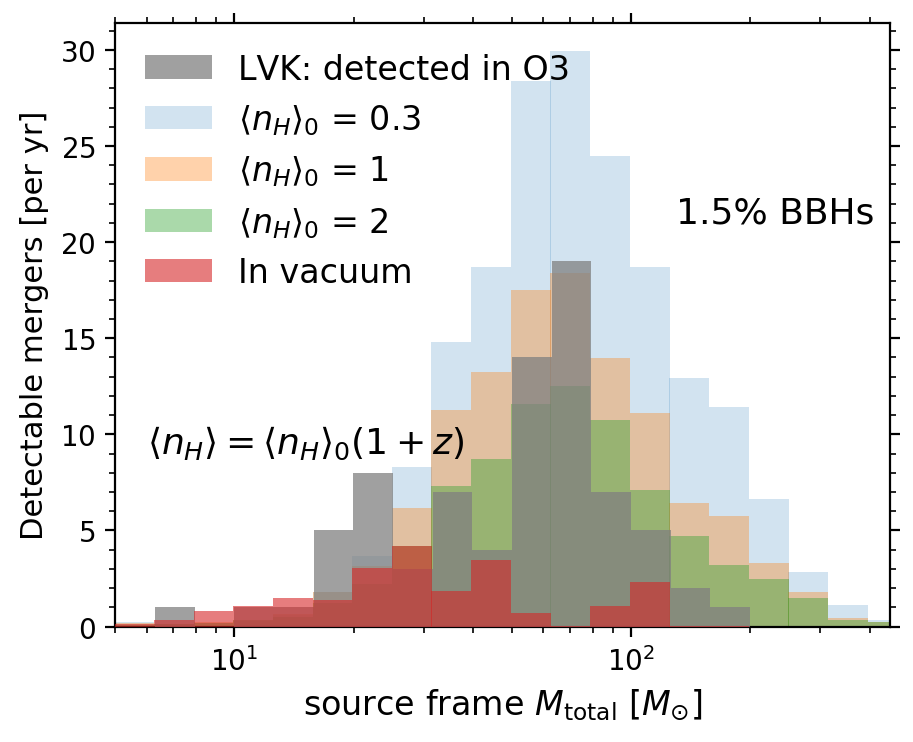}
    % \vspace{-5pt}
    \includegraphics[width = 1\linewidth]{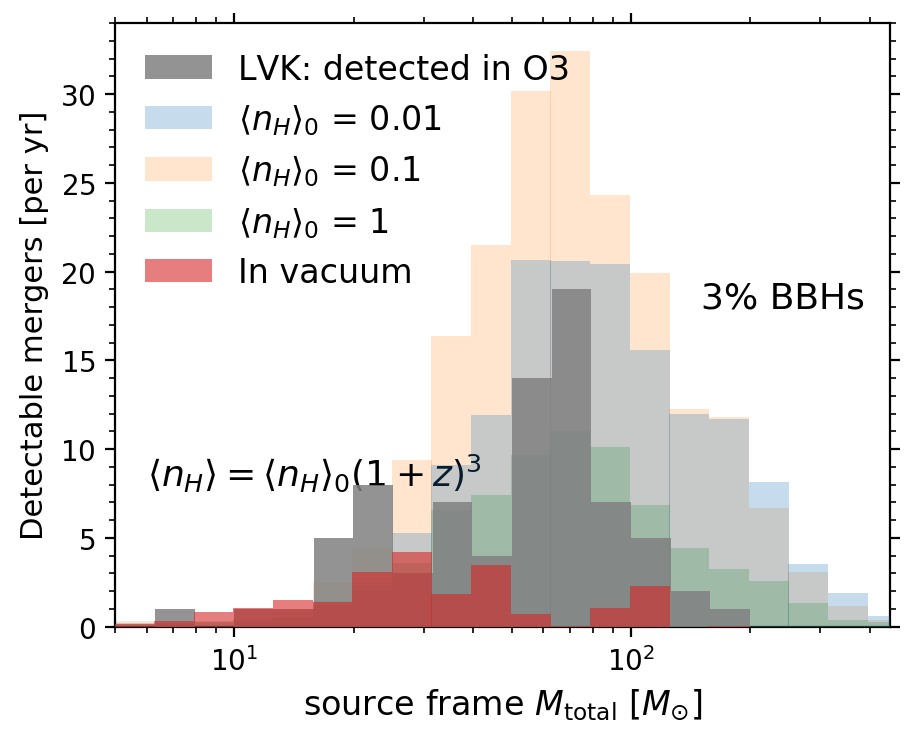}
    \vspace{-10pt}
    \caption{The detectable number of mergers on Earth in a year (as a function of the BBH mass) assuming that all three LVK detectors work in quadrature for a full year at the third observing run (O3) sensitivity. The legend labels are similar to those in Fig.~\ref{fig: volumetric merger rate}. The grey region represents the BBH mergers observed by the LVK in O3. The top figure is for linear, while the bottom is for cubic redshift dependence of $\langle n_H \rangle$. Similar to Fig.~\ref{fig: volumetric merger rate}, we assume that only 1.5\% (top figure) and 3\% (bottom figure) BBHs have an optimal ISM environment.}
    
    \label{fig: detectable merger rate}
\end{figure}

\subsubsection{Impact on the BBH's orbital decay}

In addition to the intensification of gravitational wave emission resulting from an increasing mass, e.g., see Section~\ref{subsec: decay due to GW emission}, earlier, we observed that isotropic accretion at $v_\infty = 0$ further accelerates the orbital decay process (Fig.~\ref{fig: rate of orbital decay}). This is due to the accreted matter not contributing towards the BBH's angular momentum.

In the case where $v_\infty \neq 0$, the rate of change of angular momentum $\dot{L}$ of a mass accreting binary (outside of gravitational wave emission) can be estimated as 
\begin{equation}
    \dot{{L}} = \dot{m} {d} \times {v}_\infty \,,
\end{equation}
where $d$ is the distance from the binary's center of mass. Now, depending on the magnitude and sign of the cross-product term, the accretion of mass will either contribute to or remove angular momentum from the binary. 

However, as the orbital velocity of LVK black holes is much larger than a subsonic $v_\infty$, thus averaged over an orbit, the contribution from ISM accretion to $L$ would be negligible.
Similarly, as we have already discussed in Section~\ref{subsec: Impact of mass of a non-zero velocity}, since the mass accretion rate drastically drops for supersonic motion, as such, we expect a negligible change in $L$  for large values of $v_\infty$ as well.

\vspace{-5pt}
\subsection{What if $v_\infty = 0$ but the accretion flow contains angular momentum?}

It is possible that even though a BBH is comoving w.r.t. its surrounding ISM, the medium still possesses some angular momentum w.r.t. the center of mass of the BBH (in contrast to the disk around individual black holes which can extend to very small radii, e.g, Section~\ref{sec: feedback from ISM}, the circumbinary disk will likely disrupt as it reaches either of the black holes in the binary. Due to its large truncation radius, such a disk will not have significant radiation feedback but may heat the environment if it forms mini-disks around the individual black holes).
Such a medium could comprise a generic ISM and/or the circumstellar medium, the latter being the leftover matter from the earlier episode of stellar evolution (e.g., sourced from stellar winds or due to Roche Lobe overflow from the L2 Lagrangian point). 
For any such generic angular momentum possessing material to be accreted, it would need to be lowered to the location of the black hole. At this point, the specific angular momentum of the accreted fluid element would be the same as the specific angular momentum of the black hole w.r.t. the center of mass of the binary.
As such, this will not affect the separation of the binary. On the other hand, it will make the black holes more massive, thus accelerating the gravitational radiation emission-induced inspiral.  

The orbital decay of a BBH undergoing such an accretion process (assuming that the material is efficiently lowered to the location of the black hole such that Eq.~\ref{eq: Bondi Hoyle} still remains a fair estimate) can be described as (cf. Eq.~\ref{eq: linear diff eqn})
\begin{equation}
    \frac{d r^4}{d t}   + 4 \beta_i \eta_1 \eta_2^2 = 0 \,,
    % \label{eq: linear diff eqn}
\end{equation}
which gives us the merger time relation as
\begin{equation}
    \int_{t_i}^{T_{\rm tot} + t_i}  \eta_1(t) \eta_2^{2}(t) dt = T_{\rm GW} \,.
\end{equation}
Analogous to Eq.~\ref{eq: merger time}, if we assume the binary to have equal mass components, this yields
\begin{equation}
    T_{\rm tot} = \frac{1}{m_i \alpha} \left[1 - \left(\frac{1}{2m_i \alpha  T_{GW}+ 1}\right)^{1/{2}} \right] \,.
    % \label{eq: merger time}
\end{equation}
The resulting solution is illustrated in Fig.~\ref{fig: mass evolution of a BH from GW emission only} and shows that merely the growth in black hole mass (with $\left. \frac{dr}{dt} \right|_{\rm AM} = 0$) can considerably impact the merger time scale of black holes.

\vspace{-5pt}
\section{Conclusion} \label{sec: conclusion}

Many studies attempting to reproduce the LVK inferred BBH properties assume that the black holes live in a vacuum. While this assumption is justified for a short evolutionary timescale, for longer timescales of $\mathcal{O}(10^9)$ yr (typical for LVK BBHs), one might expect some deviations. This is because the BBHs are embedded in an ISM environment whose impact on the BBH could accumulate over a span of few billion years.

Assuming that for the most part, at least a few percent of BBHs in our study are immersed in a a favorable cold ISM environment, here we discussed the impact of the latter on the BBH dynamics.
We showed that if the binary as a whole comoves w.r.t. its surrounding matter, then such a medium could act as an important driver of a subset of LVK BBH mergers. 
Additionally, it could also push some high-mass BBHs in the pair-instability mass gap. 
% Another feature of ISM-accreting BBHs would be their low effective spin due to the accretion of mass with negligible angular momentum.  
% 
% 
Apart from the above assumptions, another free parameter in the present work is the time-averaged particle number density $\langle n_H \rangle$ of the surrounding ISM.  While we approximately reproduce some of the LVK observables by fine-tuning $\langle n_H \rangle$, it is unlikely that this choice would hold in a realistic setting. However, it could be possible that a certain subset of LVK's BBHs are impacted by their environment (including their dark matter surroundings), and we hope this work will shed light on the effect of such a medium on the BBH dynamics.

\vspace{-10pt}
\section*{Acknowledgements}

I am grateful to Yuri Levin and the members of the THEA seminar for their valuable feedback during a presentation of this work and Max Briel and J.J. Eldridge for helpful discussion. I am supported by the University of Auckland Doctoral Scholarship. This work utilized NeSI high performance computing facilities.

% %%%%%%%%%%%%%%%%%%%%%%%%%%%%%%%%%%%%%%%%%%%%%%%%%%

\vspace{-15pt}
\section*{Data Availability}
The \textsc{Bpass} models used here can be found at \url{https://bpass.auckland.ac.nz/}. The jupyter notebook used for making the figures along with the associated output data can be found at \url{https://github.com/SohanGhodla/Black-holes-embedded-in-an-ISM}.



\vspace{-5pt}
\bibliographystyle{mnras}
\bibliography{refs} % if your bibtex file is called example.bib

\appendix

\vspace{-5pt}
\section{Merger rate calculation} \label{sec: rate calculation}

Here, we summarise the merger rate calculation employed in this work. For background details on the employed population synthesis, we refer the reader to Appendix A in \cite{Ghodla_coupled_black_holes:2023}. For the current work, the relevant output of the \textsc{Bpass} simulation is recorded in the $\mathcal{R}_{\rm{smpl}}(m_1, m_2, \tau, Z)$ variable. It contains the merger rate density of the  BBHs per unit mass of star formation in the simulation as a function of the component formation masses ($m_1, m_2)$, the total delay time $\tau$ (i.e., time from starburst to  BBH merger) and progenitor binary star's metallicity $Z$. 

We note that the calculated sample merger rate density $\mathcal{R}_{\rm{smpl}}(m_1, m_2, \tau, Z)$ assumes that the BBHs live in vacuum, i.e., the standard \textsc{Bpass} output does not take into account the effect of the presence of an ISM on the BBH dynamics. The latter will be treated in Section~\ref{subsec: A2}.

\subsection{Source frame volumetric merger rate for BBHs in vacuum} \label{subsec: source frame rate}

% Let us assume that the total delay time from the onset of a binary star formation to its eventual gravitational radiation emission-induced merger is $\tau$. 

For a realistic cosmological population of BBHs living in vacuum, the  source frame  volumetric BBH merger rate density $\mathcal{R} (m_1, m_2, t)$ (purely due to emission of gravitational radiation) at some time $t$ due to all the previously occurred star formation can we written as 
\begin{equation}
    \begin{aligned}
    \mathcal{R} (m_1, m_2, t)  = \int_{0}^{\infty}  dZ \int_{\Tilde{t}}^{t} & d \tau_{\rm GW}   \,\, \psi(t-\tau_{\rm GW}, Z) \\
    & \times \mathcal{R}_{\rm{smpl }}(m_1, m_2, \tau_{\rm GW}, Z)  \,.
    \label{eq: R_vol as a function of metallicity}
\end{aligned}
\end{equation}
Here, $\psi(t-\tau_{\rm GW}, Z)$ is the star formation rate density at metallicity $Z$ and $\mathcal{R}_{\rm smpl} (m_1, m_2, \tau_{\rm GW}, Z)$ is the merger rate density of BBHs per unit mass of star formation with merger delay time $\tau_{\rm GW}$ at metallicity $Z$. Also, the lower limit of integration $\Tilde{t}$ marks the beginning of star formation and is taken as the time when the Universe was $\approx 200$ Myr old. The form of $\psi$ is taken from \cite{Madau_Dickinson2014} in conjunction with the analytic expression for metallicity evolution from \cite{Langer_Norman_2006}.
 % 

% \vspace{-5}
\subsection{Source frame volumetric merger rate for BBHs embedded in an ISM surrounding} \label{subsec: A2}

The merger time of a BBH embedded in an ISM surrounding can be calculated using Eq.~\ref{eq: general merger time}. However, due to its mathematically complex nature, the calculation of the corresponding source frame merger rate for such a general solution has to be done numerically. To illustrate the calculation analytically, here we assume the merger time solution for the special case when $m_{1,i} = m_{2,i} = M_i/2$ as given in Eq.~\ref{eq: merger time}. For consistency of notation, we invert and rewrite this equation in terms of the variable $\tau$ as
\begin{equation}
    \tau_{\rm GW} = \frac{[1 - (M_i \alpha \tau_{\rm tot} / 2)]^{-14} - 1}{7M_i \alpha} \,.
    % \tau_{\rm tot} = \frac{1}{m_i \alpha} \left[1 - \left(\frac{1}{14m_i \alpha  \tau_{GW}+ 1}\right)^{1/{14}} \right] \,.
    % \label{eq: merger time}
\end{equation}
In a differential form this becomes
\begin{equation}
%     d\tau_{k=0} = \frac{d \tau_{\rm tot}}{ \left[ \frac{\tau_{\rm tot}}{t_i} + 1 \right]^{-10k}} \,.
     d\tau_{\rm GW} =  \frac{d \tau_{\rm tot}}{(1 - 0.5M_i \alpha \tau_{\rm tot})^{15}} \,.
\end{equation}
For BBHs embedded in an ISM, the earlier expression $\mathcal{R}_{\rm{smpl }}(m_1, m_2, \tau_{\rm GW}, Z) d \tau_{\rm GW} $  thus transform into
\begin{align*}
    \quad \mathcal{R}_{\rm{smpl}} & \left(m_{1,i}, m_{2,i} , \frac{[1 - (M_i \alpha \tau_{\rm tot} / 2)]^{-14} - 1}{7M_i \alpha}, Z  \right) \\
    & \quad \quad \times (1 - 0.5 M_i \alpha \tau_{\rm tot})^{-15} d \tau_{\rm tot} \,,
\end{align*}
where we have introduced the subscript ``$i$'' to indicate the initial black hole masses as the latter are now dynamical.
The delay between star formation time and the formation time of the BBH also needs to be calculated within the simulation. For the sake of demonstration, here we set this to zero, i.e.,  the BBH formation time is simply $t_i = t - \tau_{\rm tot}$.
Thus, the variant of Eq.~\ref{eq: R_vol as a function of metallicity} for BBHs embedded in an ISM surrounding takes the form
\begin{equation}
    \begin{aligned}
     & \mathcal{R}(m_{1,f}, m_{2,f}, t) = \int_{0}^{\infty} dZ \int_{\Tilde{t}}^{t} d \tau_{\rm tot} \,\, (1 - 0.5 M_i \alpha \tau_{\rm tot})^{-15}   \,\, \\ & \times \psi(t-\tau_{\rm tot}, Z) \cdot \mathcal{R}_{\rm{smpl}} \left(m_{1,i}, m_{2,i}, \frac{[1 - (M_i \alpha \tau_{\rm tot} / 2)]^{-14} - 1}{7 M_i \alpha}, Z  \right) \\
     & \times \delta(m_1(\tau_{\rm tot}) - m_{1,f}) \cdot \delta(m_2(\tau_{\rm tot}) - m_{2,f}) \,.
\end{aligned}
\end{equation}
Above, the subscript ``$f$'' represents the final masses (i.e., just before merger) and the $\delta$ symbols are the Dirac delta functions. Additionally, the value of $m(\tau_{\rm tot})$ can be calculated using Eq.~\ref{eq: mass evolution}. We note that although the function $\mathcal{R}_{\rm smpl}$ remains the same, its domain spanned by the integral in above equation has changed and now allows for mergers with much larger delay time.

\subsection{Detectable merger rate}  \label{subsec: intrinsic rate}

The maximum possible rate observed on Earth (i.e., at infinite detector sensitivity, also known as intrinsic rate) from the above calculated $\mathcal{R}$ can be computed as
\begin{equation}
    \mathcal{R}_{\rm{intr}} (m_1, m_2) =\int_{0}^{\Tilde{z}} d V(z) \,\, \frac{\mathcal{R} (m_1, m_2, t_{z})}{1 + z}  \;\; [{\rm yr}^{-1}] \,.
    \label{eq: intrinsic rate}
\end{equation}
We set $\Tilde{z}$, corresponding to the time $\Tilde{t}$ in Eq. \ref{eq: R_vol as a function of metallicity}. The $1/(1+z)$ factor scales the source time to observer time (i.e. $t$ at $z = 0$) and $dV$ is a differential comoving spatial volume element:
\begin{equation}
    d V(z) =\frac{4 \pi c}{H_{0}} \frac{D_{c}^{2} (z)}{E(z)} d z \,.
\end{equation}
Also $E(z) =\sqrt{\Omega_{M}(1+z)^{3} + \Omega_{\Lambda}}$, $H_0$ is today's value of Hubble's constant and 
\begin{equation}
    D_{c}(z) =\frac{c}{H_{0}} \int_{0}^{z} \frac{d z^{\prime}}{E\left(z^{\prime}\right)} 
\end{equation}
is the comoving distance.
It is assumed that by the time the BBHs form, we are well into the matter-dominated epoch in a flat $\Lambda$CDM cosmology with $\Omega_M = 1-\Omega_\Lambda, \Omega_\Lambda = 0.6925$ and $H_0 = 67.9$ km s$^{-1}$Mpc$^{-1}$ \citep{Planck_Collaboration_2016}. To calculate the LVK detectable rate from the above-described intrinsic rate, we follow the methodology described in Appendix A5 in \cite{Ghodla_coupled_black_holes:2023}. This assumes that all three LVK detectors work in quadrature for a full year at O3 sensitivity \citep{LVK}.

\section{Additional figures}

\begin{figure}
    \centering
    % \vspace{-5pt}
    \includegraphics[width = 1\linewidth]{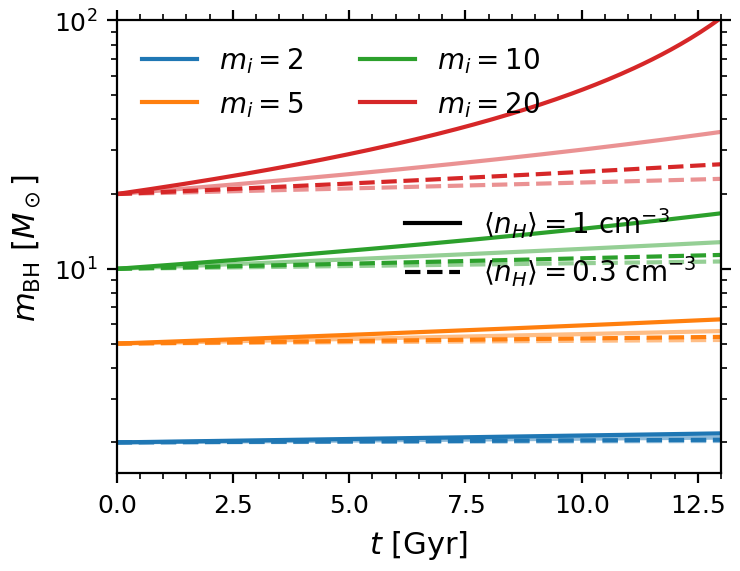}
    \vspace{-5pt}
    \caption{Same as Fig.~\ref{fig: mass evolution of a BH} but assuming $T_\infty = 200$ K (darker lines) and $T_\infty = 300$ K lighter lines.}
    \label{fig: mass evolution of a BH at larger T}
\end{figure}

\begin{figure}
    \centering
    % \vspace{-15pt}
    \includegraphics[width = 1\linewidth]{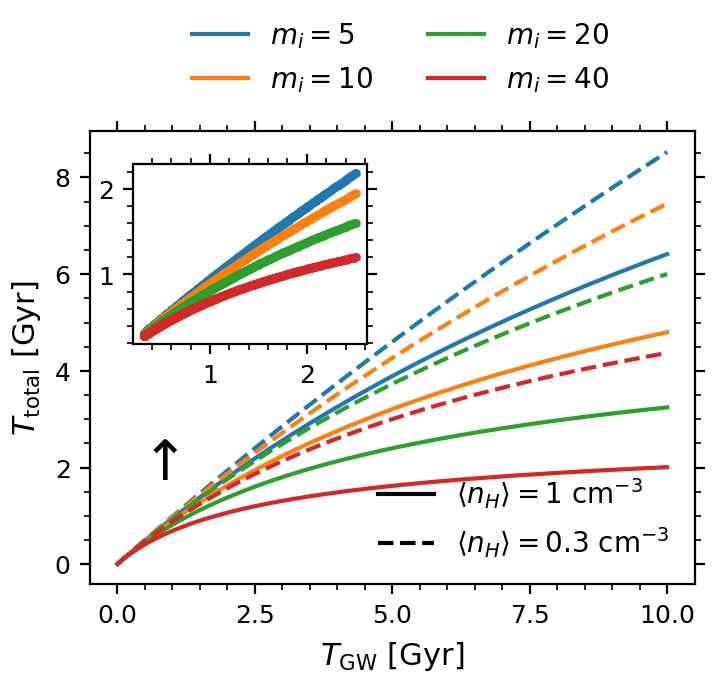}
    \vspace{-5pt}
    \caption{Same as Fig.~\ref{fig: Merger time} but now assuming that the accreted mass contains the same amount of specific angular momentum as possessed by the black hole w.r.t. the center of mass of the binary.}
    \label{fig: mass evolution of a BH from GW emission only}
\end{figure}

%%%%%%%%%%%%%%%%%%%%%%%%%%%%%%%%%%%%%%%%%%%%%%%%%%

% Don't change these lines
\bsp	% typesetting comment
\label{lastpage}
\end{document}